\documentclass[10pt]{article}
\usepackage[english]{babel}
\usepackage[dvips]{graphicx} 
\usepackage{amssymb,amsmath}

\newcommand{\zetab}{\mbox{\boldmath$\zeta$}}
\newcommand{\rl}{r_{\Lambda}}
\newcommand{\rs}{r_{\rm s}}
\newcommand{\lp}{\left(}
\newcommand{\rp}{\right)}
\newcommand{\dd}{{\rm d}}

\begin{document}

\title{Scales Set by the Cosmological Constant}
\author{
Andr\'{e}s Balaguera-Antol\'{\i}nez\footnote{e-mail: {\tt a-balagu@uniandes.edu.co}}~,
Christian G. B\"ohmer\footnote{e-mail: {\tt boehmer@hep.itp.tuwien.ac.at}}~,\\ 
and Marek Nowakowski\footnote{e-mail: {\tt mnowakos@uniandes.edu.co}} \\[1ex]
\footnotemark[1]~\footnotemark[3]
Departamento de F\'{\i}sica, Universidad de los Andes,\\ 
A.A. 4976, Bogot\'a, D.C., Colombia.\\[1ex]
\footnotemark[2]
ASGBG/CIU, Department of Mathematics, Apartado Postal C-600,\\
University of Zacatecas (UAZ), Zacatecas, Zac 98060, Mexico.}
\date{}
\maketitle

\begin{abstract}
The cosmological constant, $\Lambda$, sets certain scales important 
in cosmology. We show that $\Lambda$ in conjunction with other 
parameters like the Schwarzschild radius leads to scales relevant not only 
for cosmological but also for astrophysical applications. Of special interest 
is the extension of orbits and velocity of test particles traveling over Mpc distances. We will 
show that there exists a lower and an upper cut-off on the possible  
velocities of test particles. For a test body moving in a central
gravitational field $\Lambda$ enforces a maximal value of the
angular momentum if we insist on bound orbits of the test body which 
move at a distance larger than the Schwarzschild radius.
\end{abstract}

\section{Introduction}
Any fundamental constant of physics sets certain scales, either by 
itself or in conjunction with other constants. The latter
can be also of fundamental nature or specific constants entering a given 
physical problem. One such an example is the Planck mass or length 
which is a combination of fundamental constants, another the Schwarzschild radius
in which the mass of an object enters. These scales often govern the global behavior
of the physical system and determine the orders of magnitude of an observable 
of such a system. It is therefore of some importance  to examine the scales 
dictated by the existence of a positive cosmological constant $\Lambda$ in 
different contexts (an early example in connection with Dirac's large numbers 
which can even be considered as a prediction for $\Lambda$ can be found 
in \cite{zeld}). The physical interest is that the cause for the recently 
discovered acceleration of the universe could be a positive cosmological 
constant \cite{accel,accel2}.

Such a constant sets scales thought to be of importance only in the full cosmological framework. 
This, however, is not always the case and $\Lambda$ has some relevance 
also in astrophysical problems. The first example of this kind is given by 
the density scale set by $\Lambda$. Defining the vacuum density $\rho_{\rm vac}$ 
by $\Lambda=8\pi G\rho_{\rm vac}$, using the present value 
$\rho_{\rm vac}\approx 0.7\rho_{\rm crit}$ and imposing gravitational 
equilibrium via the virial equation with $\Lambda$ \cite{wang, barrow, ab}, we obtain for configurations with constant 
density $\rho$~\cite{we1,we12, weP}, 
\begin{equation}
      \label{i1}
      \rho\ge A \rho_{\rm vac},
\end{equation}
where $A$ is determined by the geometry of the astrophysical object under 
consideration. For spherically symmetric objects, $A=2$ such that 
$\rho\ge 1.4\rho_{\rm crit}$. If large objects (clusters, super clusters) 
have a density of at least the order of magnitude of $\rho_{\rm crit}$
i.e. $\rho>{\cal O}(\rho_{\rm crit})$, this gives a limit on 
$\rho_{\rm vac}$ in the form $\rho_{\rm vac}\leq 0.5 {\cal O}(\rho_{\rm crit})$ 
which comes very close to the present value. More interestingly, for non-spherical 
objects $A\approx (a/b)^{n}$ where $a$ and $b$ are two length scales such 
that $(a/b)>1$. For $n \geq 3$ (this is possible for ellipsoids) and $(a/b)\gg 1$, 
the inequality becomes stringent and of astrophysical relevance~\cite{we1,we12}. 
This can even be of relevance for low density galaxies, limiting either 
their `flatness' (essentially the ratio $a/b$) or their density. 
For spherically symmetric objects, inequality (\ref{i1}) emerges also 
from other equilibrium concepts, like the hydrostatic equilibrium~\cite{christian,we2}. 
The spherical collapse model in different dark energy cosmologies
was investigated by~\cite{mota}.

\section{Fundamental cosmological scales}

The actual value $\rho_{\rm vac}\approx 0.7 \rho_{\rm crit}$ can 
be regarded as a coincidence problem. As $\rho_{\rm crit}$ is epoch-dependent, 
we can put forward the question as to why we live right now in a universe 
in which $\rho_{\rm vac}={\rm const}\sim \rho_{\rm crit}(t_{0})$. 
One can reformulate the coincidence in terms of length scales. 
In the context of the Newtonian Limit, it has been noted in~\cite{we2} that 
\begin{equation}
      \label{lsca}
      r_{\Lambda}=\frac{1}{\sqrt{\Lambda}}=
      \frac{1}{\sqrt{3}}\left(\frac{\rho_{\rm vac}}{\rho_{\rm crit}}\right)^{-1/2}
      H_0^{-1}=2.95\times 10^{3}h^{-1}_{70} \ \rm Mpc,
\end{equation}
where $h_{70}$ is defined by the value of the Hubble parameter 
$H_{0}=70h_{70}$ kms$^{-1}$Mpc$^{-1}=100h_{0}$ kms$^{-1}$Mpc$^{-1}$, 
with the present value given by $h_{70}=1.0\pm 0.15$ \cite{cosmobook}.
Equation (\ref{lsca}) sets a cosmological scale which is about 
the radius of the visible universe. This coincidence becomes even 
stronger by noticing that the weak field approximation 
of Einstein's equations with $\Lambda$ is valid within the range \cite{we2}
\begin{equation}
      \sqrt{6}r_{\Lambda}\gg r \gg r_{s}\equiv G M, \quad
      M \ll M_{\Lambda}\equiv \frac{2\sqrt{2}}{3}\frac{1}{G\sqrt{\Lambda}}=4.8
      \times 10^{22}h^{-1}_{70}M_{\odot}\lp\frac{\rho_{\rm crit}}
      {\rho_{\rm vac}}\rp^{1/2}.
      \label{1}
\end{equation}
As the inequality (\ref{i1}) has a connection to the hydrostatic equilibrium, 
the scales appearing in (\ref{1}) in connection with the Newtonian limit also 
emerge in the hydrostatic equilibrium framework. Indeed, it follows from the 
so-called Buchdahl \cite{bh, buchdahl,buchdahl2} inequalities that
\begin{equation} 
      \label{B}
      r\le\sqrt{\frac{4}{3}} r_{\Lambda}, \qquad
      M\le\sqrt{\frac{2}{3}}M_{\Lambda},
\end{equation}
which sets absolute scales on the validity of the hydrostatic equilibrium concept.
The second inequalities in (\ref{1}) and (\ref{B}) bring a new cosmological scale 
of cosmological orders of magnitude close to the mass of the universe. Hence, 
the coincidence problem can be formulated as a question as to why are we living in 
a universe whose length and mass scales appear as restrictions in the weak 
field approximation and hydrostatic equilibrium. Below we shall show that 
$\Lambda$ in combination with other scales leads also to scales relevant for 
astrophysics.

\section{Extension for bound orbits}

Consider the motion of test particles in a spherically symmetric and static 
vacuum space-time in the presence of the cosmological constant. 
The Schwarzschild-de Sitter metric takes the form 
\begin{equation}
      \label{sca1}
      \dd s^{2}=-e^{\nu(r)}\dd t^{2}+e^{-\nu(r)}\dd r^{2}+r^{2}\dd 
      \theta^{2}+
      r^{2}\sin^{2}\,\negmedspace \theta \dd \phi^{2},
\end{equation}
where
\begin{equation}
      \label{sca1a}
      e^{\nu(r)}=1-\frac{2\rs}{r}-\frac{r^{2}}{3\rl^{2}}. 
\end{equation}
One would suspect that the inclusion of $\Lambda$ is irrelevant in this setting. 
Note, however, that there are two  scales involved, $\rs$ and $\rl$. The 
combination of the two can lead to new results. Indeed, the equation of 
motion for a massive particle with proper time $\tau$ in the  
Schwarzschild-de Sitter metric is given by 
\begin{equation}
      \label{sca2}
      \frac{1}{2}\lp\frac{dr}{d\tau}\rp^{2}+U_{\rm eff}=
      \frac{1}{2}\lp \mathcal{E}^{2}+\frac{L^{2}\Lambda}{3} -1 \rp
      \equiv C={\rm constant},
\end{equation}
where $\mathcal{E}$ (which should not be confused with the energy) and $L$ 
are conserved quantities defined by
\begin{equation}
      \label{sca2a}
      \mathcal{E}=e^{\nu(r)}\frac{\dd t}{\dd \tau},\qquad
      L = r^2 \frac{\dd \phi}{\dd \tau}.
\end{equation}
The function $U_{\rm eff}$, defined by
\begin{equation}
      \label{sca2b}
      U_{\rm eff}(r)=-\frac{\rs}{r}-\frac{1}{6}\frac{r^{2}}{\rl ^{2}}+
      \frac{L^{2}}{2r^{2}}-\frac{\rs L^{2}}{r^{3}},
\end{equation}
is the analog of the effective potential in classical mechanics. We now 
consider radial motion, i.e.~$L=0$. From the definition of $C$ we obtain the 
inequality 
\begin{equation} 
      \label{C}
      C>-\frac{1}{2}, 
\end{equation}
which will play a crucial role later in the derivation. 
The importance of $C$ and the inequality (\ref{C}) is that like any 
other constant of motion $C$ can be calculated in terms of initial 
values $v_0$ and $r_0$. The inequality (\ref{C}) will then set a limit 
of velocity in terms of distance or vice versa. For the limiting 
value $C=-1/2$, we have $\mathcal{E}=0$ which signals an artifact of the 
Schwarzschild coordinates. From (\ref{sca1a}) and (\ref{sca2a}) we see 
that there exist some $r=r_{\star}$ which satisfies the cubic equation 
\begin{equation} 
      \label{cubic}
      y^{3}-3y+6x=0, 
\end{equation}
with 
\begin{equation}
      y=\frac{r_{\star}}{r_{\Lambda}}, \quad
      x=\frac{\rs}{\rl}=1.94\times 10^{-23}\lp\frac{M}{M_{\odot}}\rp 
      h_{70}\lp\frac{\rho_{\rm vac}}{\rho_{\rm crit}}\rp^{1/2} \ll 1.
      \label{sca2c}
\end{equation}
The cubic equation (\ref{cubic}) has two positive solutions which can be found by 
the standard Cardano formulae. To avoid cumbersome expressions, it is 
customary to parameterize the solutions
by the 
auxiliary angle $\phi$ defined by \cite{matbuch}
\begin{equation} 
      \label{cubic2}
      \cos \phi = \cos (\sigma_0 +\pi/2) \simeq -\sigma_0= 3x.
\end{equation}
Then the two positive roots can be calculated from
\begin{eqnarray} 
      y_1&=-2\cos(\phi/3 + 4\pi/3)=-2 \cos (-x +9\pi/6), \nonumber \\
      y_2 &=-2\cos(\phi/3 + 2\pi/3)=-2 \cos (-x + 5\pi/6),
      \label{cubic3}
\end{eqnarray}
which leads approximately  to 
\begin{equation} 
      \label{limit}
      r_{\star}^{(1)}=\sqrt{3}\rl\left[1-\frac{1}{2}\left(\frac{r_s}{r_{\Lambda}}\right)^2
\right] -\rs, \qquad 
      r_{\star}^{(2)}=2\rs \left[1 +\frac{4}{3}\left(\frac{\rs}{\rl}\right)^2\right].
\end{equation} 
In other words, the condition $C=-1/2$ is satisfied at the Schwarzschild radius 
and at the edge of the universe. With the same limiting value, we have 
$|U_{\rm eff}(r_{\star})|=\frac{1}{2}$. We exclude from our consideration motion with 
$|U_{\rm eff}(r)| \ge \frac{1}{2}$ 
since it corresponds to allowing the motion of test particles inside the 
Schwarzschild radius and beyond the observed universe. The latter is a 
result of the coincidence in the sense that $\rl$ sets the scale of 
the horizon of the universe. Hence, the particles are allowed to be 
at some $r$ such that  $\rs<r<\sqrt{3}\rl$ with 
$|C|<|U_{\rm eff}(r)|<\frac{1}{2}$(for negative $C$ and $U_{\rm eff}$). 
It is clear that at a certain distance, the terms $-\rs/r$ and $r^{2}/\rl^{2}$ 
will become comparable leading to a local maximum located at
\begin{equation}
      \label{sca3}
      r_{\rm max}=\lp 3\rs \rl ^{2}\rp^{1/3}=
      9.5\times 10^{-5}\lp\frac{M}{M_{\odot}}\rp^{1/3}\lp
      \frac{\rho_{\rm crit}}{\rho_{\rm vac}}\rp^{1/3} h_{70}^{-2/3} 
      \ \rm Mpc,
\end{equation}
with
\begin{equation}
      \label{sca4}
      U_{\rm eff}(r_{\rm max})=-7.51\times 10^{-16} 
      \lp\frac{M}{M_{\odot}}\rp^{2/3}\lp\frac{\rho_{\rm vac}}
      {\rho_{\rm crit}}\rp^{1/3} h_{70}^{2/3}.
\end{equation}
Beyond $r_{\rm max}$, $U_{\rm eff}$ is a continuously decreasing function,
which implies that $r_{\rm max}$ is the maximum value within which we can 
find the bound solutions for the orbits of a test body. The behavior of the 
effective potential at moderate distances is governed by the scales $\rs$ and $L$. 
At large distances, $\rs$ and $\rl$ take over until finally at cosmological 
distances only $\rl$ is of relevance. Therefore, in a good approximation, 
(\ref{sca3}) and (\ref{sca4}) are also valid for most of the cases of 
moderate $L\neq 0$.

Consider now the following chain of matter conglomeration of astrophysical objects: 
the smallest are star clusters (globular and open) with stars as members 
($M=M_{\odot}$) and a mass of $10^{6}M_{\odot}$. We proceed to galaxies and 
galactic clusters. Within this chain, we find for $r_{\rm max}$ in Table 1 the  
values of $r_{\rm max}$ as a function of mass.

\begin{table}
\begin{center}
\begin{tabular}{|c|c|c|c|}\hline
\textbf{Component } &$M/M_{\odot}$  & $r_{\rm max}/\alpha $ [pc] & \textbf{Structures} \\ \hline
Stars                  &$1$            & $75$                       &Globular cluster \\ \hline
Globular cluster      &$10^{6}$       & $7.5\times 10^{3}$         & Galaxy  \\ \hline
Galaxy                &$10^{11}$      & $3.5\times 10^{5}$         & Galactic cluster\\ \hline
Galactic cluster      &$10^{13}$      & $1.6\times 10^{6}$         & Superclusters\\ \hline
\end{tabular}
\end{center}
\caption[The scales set by the cosmological constant.]{$r_{\rm max}$ in as a function of mass with $\alpha=h_{0}^{-2/3}\lp\rho_{\rm vac}/\rho_{\rm crit}\rp^{-1/3}$}
\label{scal}
\end{table}

The last value in the first line of Table 1 is of the order of magnitude of the tidal radius 
of globular clusters \cite{tremaine}. The second line agrees with the 
extension of an average galaxy. The next two values are about the size 
of a galaxy  cluster. The value $10^{13}M_{\odot}$ corresponds to a giant 
elliptic galaxy encountered often at the center of the clusters. Hence, 
$\rl$ in combination with $\rs$ gives us a surprisingly accurate and 
natural astrophysical scale. The combination $r_{\rm max}=(3\rs \rl^{2})^{1/3}$ 
from which these scales where calculated is not an arbitrary combination 
with length dimension, but it is the distance beyond which we cannot find bound 
orbits for a test body moving in a central gravitational field of an object with mass $M$.
This, however, still does not explain why the scales in Table 1 of multi-body
objects come so close the real values encountered in nature. We offer here
an attempt of an explanation. Assuming that any two members of the astrophysical
conglomeration are in an attractive gravitational interaction, this would hold also for members
at the edge of such a conglomeration. It is then not unreasonable that a result from a two
body problem which sets the scale of the extension of the two body bound system, 
is of relevance also in a multi-body environment provided the latter is not too dense. 
This is to say, such a scale does not change drastically when going 
from two body bound system to few or multi-body conglomeration.    
We would then expect that $r_{\rm max}$ sets a relevant 
astrophysical scale for large objects as demonstrated in Table 1.
Of course, we are talking here about scales neglecting dynamical aspects 
of many body interactions, but tentatively, $r_{\rm max}$ is roughly the 
scale to be set for bound systems. Indeed, the agreement of the result 
in Table 1 with values encountered in nature is striking. The 
second line in Table 1 requires some comments. Unlike the other 
three, where the argument $M$ of $r_{\rm max}$ has been taken 
to be the mass of the average members of one type of an astrophysical object, 
it might appear unjustified to take the mass of the globular cluster 
to obtain the extension of the galaxy. However, in view of the fact 
that globular clusters are very old objects and are thought to be of 
importance in the formation of the galaxy, this choice seems justified. 
Indeed, with $\Lambda >0$, our result strengthens the belief that globular 
clusters are relics of formation of galaxies. For instance, 
$r_{\rm max}$ for open star clusters with a mass $M=250M_{\odot}$ 
is only 0.5 kpc. As as side remark we note that $M=10^6 M_{\odot}$ is also
the mass of the central black hole in our galaxy. The extension
of bound orbits around this black hole is limited to be roughly $10$ kpc.

It makes sense the compare the above results which refer to the actual
extension of an astrophysical object to the virial radius ${\cal R}_{\rm vir}$.
The cosmological constant sets in the context of virial equations a maximally
possible radius of a virialized object which formally looks similar
to $r_{\rm max}$. The scalar virial equation with $\Lambda$ reads \cite{we1,we12,Jackson}
\begin{equation} \label{virial}
2{\cal K} -|{\cal W}| +\frac{1}{3}\Lambda {\cal I}=0,
\end{equation}
where ${\cal K}$, ${\cal W}$ and ${\cal I}$ are the traces of the kinetic,
potential and inertia tensor, respectively. For a spherical astrophysical
object with mass ${\cal M}$ (note that this is not the average
mass $M$ of the members of this object) and radius ${\cal R}$ we obtain
\begin{equation} \label{virial2}
{\cal R}^3 +\left( 10 G {\cal K}r^2_{\Lambda}/R_s \right){\cal R} -3R_{s}r^2_{\Lambda}=0,
\end{equation}
where we have used $\mathcal{I}=\frac{3}{5}{\cal M} \mathcal{R}^{2}$, $|\mathcal{W}|
=\frac{3}{5}\frac{G{\cal M}^{2}}{\mathcal{R}}$ and with $R_{s} \equiv G{\cal M}$. 
If ${\cal K}=0$ we get the largest possible virial radius given by
\begin{equation} \label{virial3}
{\cal R}^{\rm max}_{\rm vir} =\left(3R_sr^2_{\Lambda}\right)^{1/3}.
\end{equation}
Although formally this result looks similar to $r_{\rm max}$, both scales
signify physically a different concept. Whereas $r_{\rm max}$ is the scale
for the actual extension of a bound system (at least in a two body setting 
and possibly also beyond as explained above), ${\cal R}_{\rm vir}^{\rm max}$ 
is the maximally possible radius if $\Lambda > 0$ i.e. it is the virial 
radius when the average velocity of the components of this system is zero
on account of ${\cal K} =0$ and ${\cal K} \propto \langle v^2\rangle $. 
It is clear from Table 1 that using ${\cal M}$ instead of $M$ 
(this is the difference between $r_{\rm max}$ and ${\cal R}_{\rm vir}^{\rm max}$)
will result in virial radii which are too large. But this only means 
that ${\cal K}=0$ is not a good assumption.

\section{Velocity bounds}

The above scales of extension of astrophysical bodies still 
do not complete the full picture. We can also gain 
some insight into the velocity of particles traveling over Mpc distances. 
Particles starting beyond the astrophysical scale $r_{\rm max}$ with 
initial values such that 
\begin{equation}
      \label{sca5}
      C<U_{\rm eff}(r_{\rm max})<0,
\end{equation}
do not reach our galaxy due to the potential barrier beyond $r_{\rm max}$. 
To determine what this exactly means, let us consider a test particle with 
radial motion and calculate $C$ in terms of the initial (coordinate) 
velocity $v_{0}$ and radial coordinate $r_{0}>r_{\rm max}$ as  
\begin{equation}
      \label{sca6}
      C=C(v_{0},r_{0})=\frac{v_{0}^{2}+2e^{2\nu(r_{0})}U_{\rm eff}(r_{0})}
      {2\lp e^{2\nu(r_{0})} -v_{0}^{2}\rp}.
\end{equation}
Combining this with inequality (\ref{sca5}) we can solve for the initial velocity. 
The condition  $C<0$ can be achieved in two different situations, namely, 
the numerator is positive and the denominator is negative (case (i)) and 
vice versa (case (ii)). In analyzing this situation the power and 
importance of the inequality (\ref{C}) i.e. $C > -1/2$ together 
with $U_{\rm eff}(r) > -1/2$ will become apparent. Let us first 
establish a hierarchy of some quantities involved in the study. Defining
\begin{equation}
      \label{sca7}
      \Xi(r_0) := 2e^{2\nu(r_{0})} 
      \left[\frac{|U_{\rm eff}(r_0)|-|U_{\rm eff}(r_{\rm max})|}
      {1-2|U_{\rm eff}(r_{\rm max})|} \right],
\end{equation}
we note that $\Xi(r_0) < e^{2\nu(r_0)}$ due to $|U_{\rm eff}(r_0)| \le 1/2$. 
It is also easy to start from the obvious inequality
$2e^{2\nu(r_0)}|U_{\rm eff}(r_{\rm max})|(2|U_{\rm eff}(r_0)|-1) < 0$ 
to arrive at $\Xi(r_0) < 2e^{2\nu(r_0)}|U_{\rm eff}(r_0)|<e^{2\nu(r_0)}$.

In the case (i) ($v_0^2 > e^{2\nu(r_0)}> 2|U_{\rm eff}(r_0)|e^{2\nu(r_0)}$)  
we obtain a solution for the initial velocity from (\ref{sca5})
\begin{equation}
      \label{sca8}
      \Xi(r_{0})<2e^{2\nu(r_0)}|U_{\rm eff}(r_0)| < e^{2\nu(r_{0})}<v_{0}^{2}.
\end{equation}
However, the condition $|C(v_0, r_0)| \le 1/2$ implies in this case 
$|U_{\rm eff}(r_0)| \ge 1/2$ which would require that our $r_0$ is 
bigger than $r_{\star}^{(1)}=\sqrt{3}r_{\Lambda}$ or smaller than
$r_{\star}^{(2)}=2r_s$, following equation (\ref{limit}). 
Therefore it might appear that the case (i) is irrelevant. To see that 
this is not so, it is best to consider the problem at hand without 
the potential barrier posed by the cosmological constant i.e.~without 
equation (\ref{sca5}). If satisfying  
$r_{\star}^{(2)} < r_0 < r_{\star}^{(1)}$ (equivalent to $|U_{\rm eff}(r_0)|\le 1/2$) 
we insist on an initial value $v_0$ such that $v_0 > e^{\nu(r_0)}$ for
an arbitrary $r_0$, we will violate the fundamental inequality (\ref{C}) 
i.e. $C > -1/2$. For, choosing $C>0$ we automatically get
$v_0^2 < 2e^{2\nu(r_0)}|U_{\rm eff}(r_0)|< e^{2\nu(r_0)}$ violating 
the assumption. If $C< 0$, then $|C| \le 1/2$ leads to
$|U_{\rm eff}(r_0)| \ge 1/2$  violating again one of the restrictions.  
Hence for any $r_0$ which we parameterize as $r_0=\zeta r_{\rm max}$, 
the quantity
\begin{eqnarray}
      \label{sca10}
      v_{\rm max}(\zeta)&=&e^{\nu(r_0)}=1-\lp\frac{8}{3}\rp^{1/3}x^{2/3}f(\zeta)\\ \nonumber
      &=&1-1\times 10^{-15} \lp\frac{M}{M_{\odot}}\rp^{2/3} h_{70}^{2/3}\lp
      \frac{\rho_{\rm crit}}{\rho_{\rm vac}}\rp^{-1/3}f(\zeta),
\end{eqnarray} 
where $f(\zeta)=(2+\zeta^{3})/2 \zeta$, represents the maximal value 
of the initial velocity which we can choose at any $r_0$. This statement 
is independent of $\Lambda$ and hinges solely on the conditions $C \ge -1/2$ 
and $U_{\rm eff}(r) \ge -1/2$. As long as we are above the Schwarzschild horizon
and below the cosmological one, this statement is not an artifact of the 
coordinate system as the condition (\ref{C}) is valid throughout the 
coordinate system. At least for the simple case of a one-body problem in 
a central gravitational field, we can then say that there is an 
energy cut-off due to the inequality (\ref{C}). It is clear that the restriction 
(\ref{C}) does not admit all possible velocities $v_0$. As said above, 
the existence of $v_{\rm max}$ is independent of $\Lambda$. But $\Lambda$ 
does have an effect on the functional form of $v_{\rm max}$. It is easy to 
see from (\ref{sca10}) that $v_{\rm max}$ and 
$\gamma(\zeta)=(1-v_{\rm max}(\zeta)^{2})^{-1/2}$ have a local 
maximum due to the $\zeta^2$ term in $f(\zeta)$ which comes from 
the contribution of the cosmological constant to $U_{\rm eff}$. 
The local maximum occurs at $\zeta=1$. In Figure 1 we plot the relativistic 
factor $\gamma(\zeta)$ versus $\zeta$ with and without $\Lambda$. 
\begin{figure}
\begin{center}
\includegraphics[width=8cm,height=7cm]{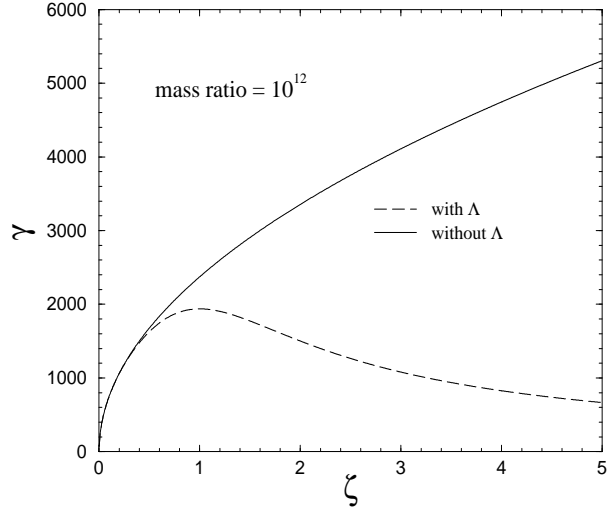}
\end{center}
\caption[The relativistic factor $\gamma=(1-v_{\text{max}}^2)^{-1/2}$]{The relativistic factor $\gamma=(1-v_{\text{max}}^2)^{-1/2}$ as a function of $\zetab=r_0/r_{\text{max}}$ with and without $\Lambda$. The mass ratio $M/M_{\odot}$ is $10^{12}$.}
\label{fig2}
\end{figure}

Let us now come to the effect of the cosmological constant i.e. to the 
task of resolving (\ref{sca5}) with respect to $v_0^2$ according to the 
case (ii): $v_0^2 < 2|U_{\rm eff}(r_0)|e^{2\nu(r_0)} < e^{2\nu(r_0)}$. 
From (\ref{sca5}) we obtain 
\begin{equation}
      \label{sca9}
      v_{0}^{2}<\Xi(r_{0})<2e^{2\nu(r_0)}|U_{\rm eff}(r_0)| < e^{2\nu(r_{0})},
\end{equation}
which means that particles whose initial velocity is smaller than 
\begin{eqnarray}
 \label{align}
      v_{\rm min}(\zeta >1 )&=\sqrt{\Xi(r_{0})}=
      x^{1/3}\lp\frac{1}{3}\rp^{1/6}\left[1-\lp\frac{8}{3}\rp^{1/3}x^{2/3}f(\zeta)\right]
      \sqrt{\frac{2f(\zeta)-3}{1-(3x)^{2/3}}}
      \\ \nonumber
      &\approx 2.24\times 10^{-8} \sqrt{2f(\zeta)-3}\lp\frac{M}{M_{\odot}}\rp^{1/3} 
      h_{70}^{1/3}\lp\frac{\rho_{\rm crit}}{\rho_{\rm vac}}\rp^{-1/6},
\end{eqnarray}
do not reach the central object with mass $M$ due to the potential barrier 
caused by $\Lambda$ (where we have used $x\ll1$ for the second line). Since we
start beyond $r_{\rm max}$ it is necessary to impose $\zeta >1 $. 
The existence of $v_{\rm min}$ means among other things, that 
cosmic rays have also a lower cut-off on energy when traveling over 
Mpc distances in space-times with a positive 
$\Lambda$. Note that the results applies equally to any kind test particle:
asteroids, comets, rockets etc. For $\zeta\to 1$ (again with $x\ll 1$), 
$v_{\rm min}(\zeta)$ grows with the distance as 
$v_{\rm min}(\zeta)\propto x^{1/3}(\zeta -1)$ and is 
evidently small. But this is a relative statement since small velocities 
for cosmic rays are not necessarily small for macroscopic objects. 
To see how $v_{\rm min}$ behaves for large $r_0$ i.e. large $\zeta$, 
we plotted in Figure 2 this quantity together with $v_{\rm max}$ 
against $\zeta$. Almost for all reasonable values of $M$, $v_{\rm min}$ 
has a local maximum at a $\zeta$ value corresponding to $r_{\Lambda}$ 
with $v_{\rm min}(r_{\Lambda})=\sqrt{4/27}=0.385$. Clearly this maximum 
is presently at the edge of the universe in which we live. This is connected 
to the coincidence problem in the sense that at later epochs, when the 
universe is much larger, this will not be the case anymore. The region 
between $v_{\rm max}$ and $v_{\rm min}$ is the region of allowed 
velocities such that the test particle can reach the central object with mass $M$.
\begin{figure}
\begin{center}
\includegraphics[width=7cm,height=6cm]{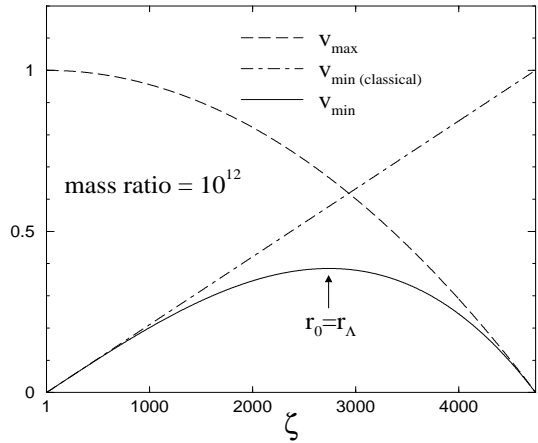}
\end{center}
\caption{$v_{\rm max}$ and $v_{\rm min}$ versus $\zeta$ for large 
         distances $r_0 \ge r_{\rm max}$ i.e. $\zeta \ge 1$. 
         $v_{\rm min}(\rm classical)$ is the velocity emerging in 
         classical non-relativistic mechanics with $\Lambda$. 
         The arrow indicates that the local maximum occurs at $r_{\Lambda}$.}
\end{figure}
Summarizing the above, there are three distinguished regions of possible 
initial velocity. Due to the restriction $C \ge -1/2$ the region 
$v_0^2 >e^{2\nu(r_0)}$ is not accessible. Only, if the initial velocity 
lies in the range $\sqrt{\Xi(r_{0})}<v_{0}<e^{\nu(r_{0})}$, they can be 
detected. If $\sqrt{\Xi(r_{0})}> v_{0}$ the test particle does not reach 
the central object. In our estimate, we neglect the fact that many  
galaxies are of spiral type and not spherically symmetric. This however, 
does not play a role, since $r_{0}$ is of the order of magnitude of Mpc 
and we consider here the problem of particles transversing Mpc distances. 
We do not consider here the details of what happens close to the galaxy.

For several reasons it is instructive to compare the above results with 
ones we would obtain in non-relativistic mechanics working in the Newtonian 
limit with $\Lambda$. Let us first point out the similarities between the 
general relativistic approach and the one in classical mechanics. For $L=0$, 
the effective potential is the same in both cases which results in the 
same $r_{\rm max}$ and $U_{\rm eff}(r_{\rm max})$. In classical 
mechanics the equation of motion is similar, but not equivalent to (\ref{sca2}). 
The constant $C$ has in this case the meaning of (non-relativistic) energy $E$ 
over mass and the affine parameter is time. Imposing the condition 
$E/m < U_{\rm eff}(r_{\rm max})$  corresponding to (\ref{sca5}) gives 
us $v^2_{\rm min}=2(|U_{\rm eff}(r_0)| -|U_{\rm eff}(r_{\rm max})|)$ 
which grows in an unlimited way unlike the general relativistic expression (see Figure 2). 
The difference to general relativity is twofold. First, $C$ is not energy. 
In addition we have the constraint $C > -1/2$ which is absent in 
classical mechanics. Of course, by hand we could restrict the velocity 
to account for the fact that in classical mechanics we are in the 
non-relativistic realm. But this would simply signal an inadequacy of 
classical mechanics. Secondly, formally the constant $C$ is also included 
in the derivative with respect to the affine parameter which makes the 
relation between $C$ and $v^2$ nonlinear (see equation (\ref{sca6})) 
unlike the relation between $E$ and $v^2$ in classical mechanics. As far
as this is concerned, our result on the existence of $v_{\rm max}$ is a 
genuine general relativistic result. Similarly the largest value of 
$v_{\rm min} \sim 0.4$ is due to general relativistic effect.

\section{Maximal angular momentum}

To discuss the full case of non-zero angular momentum i.e. $L\neq 0$ requires more
complicated calculations as we end up with higher order polynomial
equations. We will therefore discuss here only one result which can be 
obtained analytically. This result concerns the maximally possible 
$L$, due to $\Lambda$, such that bound orbits beyond the Schwarzschild 
radius are still possible. In case $\Lambda=0$, it is known that we require \cite{carroll}
\begin{equation} 
\label{l1}
L > L_{\rm min}\equiv 2\sqrt{3}r_s,
\end{equation}
to ensure bound orbits. If $r_{\Lambda} \gg L \gg r_s$ this will not change much
in the presence of the cosmological constant. 
Qualitatively, for $L \neq 0$ and $\Lambda >0$ the functional form of
$U_{\rm eff}$ is as follows. Depending on the values of the parameters, 
there can be three local extrema: a maximum close to $r_s$, a minimum 
followed by a maximum due to $\Lambda$.
We will show now that due to $\Lambda$ there exist also a maximal 
critical value $L_{\rm max}$ such that bound orbits with strictly $r > r_s$
are only possible if
\begin{equation} \label{l2}
      L < L_{\rm max}.
\end{equation}
$L_{\rm max}$ exists if the local maximum of $U_{\rm eff}$ due to $\Lambda$ and the
standard local minimum where we find the bound orbits fall together 
i.e. if we demand
\begin{equation} \label{l3}
\frac{\dd U_{\rm eff}}{\dd r}=\frac{\dd ^2U_{\rm eff}}{\dd r^2}=0.
\end{equation}
Instead of the two extrema we will get a saddle point and no
bound orbits will be possible. The two equations in (\ref{l3}) 
can be written explicitly as
\begin{eqnarray} \label{l4}
r^5 &=&3\rs \rl^{2} r^{2}-3\rl^{2}L^{2}r+9\rl^{2}\rs L^{2}, \nonumber \\
r^5 &=&-6\rs \rl^{2} r^{2}+9\rl^{2}L^{2}r-36\rl^{2}\rs L^{2}.
\end{eqnarray}
These can be solved for the position of the saddle point $r=r_{\times}$ 
and $L_{\rm max}$. By equating these expressions we obtain the quadratic equation for $r_{\times}$
\begin{equation}
\label{uh}
r_{\times}^{2}-\lp\frac{4L^{2}}{3\rs}\rp r_{\times}+5L^{2}=0.
\end{equation}
For the saddle point one finds two solutions 
\begin{equation}
r_{\times}^{(1)}\approx \frac{4}{3}\frac{L^{2}}{\rs},\,\hspace{1cm} r_{\times}^{(2)}\approx \frac{15}{4}\rs,
\end{equation}
under the assumption $L\gg \rs$. Note that the second root is of the order of the Schwarzschild's radius. 
With the first root, $r^{(1)}_{\times}$,
inserted into the first equation in (\ref{l4}) we obtain a fourth order equation for $L=L_{\rm max}$ 
\begin{equation} \label{l5}
y_{\times}^4-\alpha^2y_{\times}-12\alpha^3=0,
\end{equation}
with
\begin{equation} \label{l6}
y_{\times}\equiv \left(\frac{L_{\rm max}}{r_{\Lambda}}\right)^2,\quad
\alpha \equiv \left(\frac{3r_s}{4r_{\Lambda}}\right)^2, \quad
\end{equation}
The procedure to determine the solutions of (\ref{l6}) is to find the zeros $z_1$, $z_2$ and $z_3$ of the associated 
third order equation which in our case takes the form \cite{matbuch} (interestingly, third order
polynomial equation appear also in connection with $\Lambda$  considering purely cosmological contexts \cite{felten})
\begin{equation} \label{l7}
z^3+48\alpha^3z-\alpha^4=0. 
\end{equation}
In terms of the roots of (\ref{l7}), the roots for Eq. (\ref{l6}) can be written as
\begin{eqnarray} \nonumber
y_{1}&=& \frac{1}{2}\lp\sqrt{z_{1}}+\sqrt{z_{2}}-\sqrt{z_{3}} \rp, \\ \nonumber
y_{2}&=& \frac{1}{2}\lp\sqrt{z_{1}}-\sqrt{z_{2}}+\sqrt{z_{3}} \rp, \\ \nonumber
y_{3}&=& \frac{1}{2}\lp-\sqrt{z_{1}}+\sqrt{z_{2}}+\sqrt{z_{3}} \rp, \\
y_{4}&=&  \frac{1}{2}\lp-\sqrt{z_{1}}-\sqrt{z_{2}}-\sqrt{z_{3}} \rp. 
\label{co17}
\end{eqnarray}
%Then, for instance, the combination
%\begin{equation} \label{l8}
%y_1=\frac{1}{2}\left[\sqrt{z_1} -\sqrt{z_2} +\sqrt{z_3}\right],
%\end{equation}
%is the zero of the fourth order equation.
For the zeros $z_i$ we obtain in the limit $\alpha \ll 1$ (as before, it is convenient to parameterize
the solutions by an angle $\phi$ which is here defined by $\phi={\rm sinh}^{-1}(1/128\alpha^{1/2})$):
\begin{eqnarray} \nonumber
z_1 &=& 8\alpha^{3/2}\sinh\left[\frac{1}{3}{\rm sinh}^{-1} \lp\frac{1}{128}\alpha^{-1/2}
\rp \right] \simeq \alpha^{4/3}, \\ \nonumber
z_2 &=& -4\alpha^{3/2}\sinh\left[\frac{1}{3}{\rm sinh}^{-1}\lp \frac{1}{12}
\alpha^{-3/2}\rp \right]-4i\sqrt{3}\alpha^{3/2}\cosh
\left[\frac{1}{3}{\rm sinh}^{-1}\lp\frac{1}{128}\alpha^{-1/2}\rp\right], \\ \nonumber
&\simeq &  -\alpha^{4/3}\exp\lp \frac{i}{3}\pi\rp, \\ \nonumber
z_3 &=&  -4\alpha^{3/2}\sinh\left[\frac{1}{3}{\rm sinh}^{-1}\lp \frac{1}{12}\alpha^{-3/2}\rp \right]
+4i\sqrt{3}\alpha^{3/2}\cosh\left[\frac{1}{3}{\rm sinh}^{-1}\lp\frac{1}{128}\alpha^{-1/2}\rp\right],\\
&\simeq &    -\alpha^{4/3}\exp\lp-\frac{i}{3}\pi\rp,\label{l9}
\end{eqnarray}
where we have used the fact that for large arguments $x$ we can approximate
\begin{equation} \label{l10}
{\rm sinh}^{-1}x =\ln(x+\sqrt{x^2+1})\simeq \ln(2x)\simeq {\rm cosh}^{-1}x.
\end{equation}
At the end of the lengthy calculation we get that the only real 
and positive root for the equation (\ref{l5}) is the expression for $y_{2}$ given in (\ref{co17}). We obtain
\begin{equation} \label{l11}
y_{2}=\alpha^{2/3}=\left(\frac{3r_s}{4r_{\Lambda}}\right)^{4/3},
\end{equation}
A check that (\ref{l11}) is indeed a good approximation can be done
by neglecting the $\alpha^3$ term in (\ref{l5}). Equation (\ref{l11} implies for the maximum angular momentum 
\begin{equation} \label{l12}
L_{\rm max}=\left(\frac{3}{4}\right)^{2/3}\left(r^2_sr_{\Lambda}\right)^{1/3} 
           = 1.46\times 10^{-12}\lp\frac{M}{M_{\odot}}\rp^{2/3}h_{70}^{-1/3}\lp\frac{\rho_{\rm vac}}{\rho_{\rm crit}}\rp^{-1/6}\hspace{0.2cm}{\rm Mpc},
\end{equation}
and for the position of the saddle point
\begin{equation} \label{l13}
r_{\times}=\lp\frac{3}{4}\rp^{1/3}(\rs \rl ^{2})^{1/3},
\end{equation}
which corresponds to $\approx 0.6 r_{\rm max}$. The values of $L$ in 
astronomical units for different masses are given as follows
\begin{equation}
      \label{scales2}
      \text{
      \begin{tabular}{ccc}
	$M/M_{\odot}$ && $L_{\rm max}/\tilde{\alpha} $ (A.U.) \\
	$1$ && $0.3$  \\
	$10^{6}$ && $649$  \\
	$10^{11}$ && $1.3\times 10^{6}$ \\
	$10^{13}$ && $3\times 10^{7}$\\
\end{tabular}}
\end{equation}
where $\tilde{\alpha}=h_{70}^{-1/3}(\rho_{\rm vac}/\rho_{\rm crit})^{-1/6}$.
The values in (\ref{scales2}) are not very large. Indeed, as in the 
other scales this comes out because we combine a big scale $r_{\Lambda}$ 
with a smaller scale $r_s$. Note in this context that only on dimensional grounds 
$L \propto r_0$ where $r_0$ is the initial value of the radial component 
which can make $L_{\rm max}$ relevant for astrophysical objects, even for 
the solar system. Since $L$ vanishes if the perpendicular velocity 
$v^{\perp}$ is zero, we know that $L \propto v^{\perp}_0 r_0$. Therefore
the relevance of $L_{\rm max}$ and inequality (\ref{l2}) will grow 
for relativistic velocities such that $L \sim {\cal O}(r_0)$.

Another phenomenological relevance of $L_{\rm max}$ is to use this
result in a non-relativistic equation for the trajectories i.e.
$r(\vartheta)=p_0/(1-\epsilon \cos \vartheta)$ where $p_0(L)=L^2/r_s$.
We can estimate the order of magnitude of the maximal orbit's extension
by considering now
\begin{equation} \label{l23}
(p_0)_{\rm max} \equiv p_0(L_{\rm max})=\left(\frac{3}{4}\right)^{1/3}(r_sr_{\Lambda}^2)^{1/3}
=0.63\,r_{\rm max}.
\end{equation}
This result is non-relativistic, but applies now also to non-zero angular momentum.
It is satisfactory to know that the maximal extension of the bound orbit in the case $L \neq 0$ is just
one half of $r_{\rm max}$ which applied to zero angular momentum.
This result is in agreement with the location of the saddle point.
 
\section{Conclusions}

We have shown that the cosmological constant sets several important scales. 
The cosmological scales $r_{\Lambda}$ and $M_{\Lambda}$ appear also in 
the astrophysical context as values limiting the validity of the
Newtonian limit and the hydrostatic equilibrium. 
Of course at a later stage of the expansion, $r_{\Lambda}$ 
and $M_{\Lambda}$ will just become two scales related to $\Lambda$
and not to cosmology anymore. The other scales we discussed in connection with the
cosmological constant are of astrophysical orders of magnitude. This is possible
because e.g. $r_{\Lambda}$ combines with a smaller scale like $r_s$. 
This happens in $r_{\rm max}$ which is the largest radius for which 
bound orbits are possible in case of $L=0$. Similarly in 
${\cal R}^{\rm max}_{\rm vir}$ which is the largest possible radius
for virialized objects. We have discussed the effect of $\Lambda$ on 
initial velocities when the test body is moving radially. 
The effect of the cosmological constant in the equation of motion
is to induce a 'potential barrier'  which forbids particles
with a small velocity to reach the central object. Finally, 
for $L \neq 0$ we found a maximal angular momentum due to $\Lambda$ 
by demanding bound orbits beyond the Schwarzschild radius.
Equations (\ref{l13}) and (\ref{l23}), valid for non-zero $L$, are in good agreement
with the previous findings i.e. $r_{\rm max}$.   Phenomenologically, $r_{\rm max}$ being the
maximal extension of bound orbits is of the order of magnitude of observed astrophysical objects
(see Table 1). This rather surprising result suggests that $\Lambda$ sets the scale of the
size of such objects. We emphasize here again the scale aspect since for a complete
knowledge of a trajectory many other effects would be necessary. An extension of our results
could go in the direction to include $L \neq 0$ in the velocity bounds (to make a 
comparison with observations more realistic) and to consider a two body problem instead of a
motion of a test particle.

\section*{Acknowledgments}
The work of CGB was supported by research grant BO 2530/1-1 of the
German Research Foundation (DFG).\\
MN would like to thank N.~G.~Kelkar for useful discussions.

\end{document}